\documentclass[journal]{IEEEtran}

\usepackage{xcolor,soul,framed} 

\colorlet{shadecolor}{yellow}
\usepackage[pdftex]{graphicx}
\graphicspath{{../pdf/}{../jpeg/}}
\DeclareGraphicsExtensions{.pdf,.jpeg,.png}

\usepackage{algorithm}
\usepackage{algpseudocode}
\usepackage{tabularx}
\usepackage{graphicx}
\usepackage{amsfonts}
\usepackage{caption}
\usepackage{subcaption}
\usepackage{epstopdf}
\usepackage{breqn}
\usepackage{color,soul}
\usepackage{array}
\usepackage{authblk}
\usepackage{tabularx}
\usepackage{amsmath,graphicx}
\usepackage{mathtools}
\usepackage{amsthm}
\usepackage{breqn}
\usepackage{cite}
\usepackage{epstopdf}
\usepackage{array}
\usepackage{authblk}
\usepackage{mathtools} 
\usepackage{extarrows} 
 \usepackage{booktabs}
 \usepackage{hyperref}

\usepackage[justification=centering]{caption}

\usepackage{relsize}

\newcommand{\overbar}[1]{\mkern1.5mu\overline{\mkern-1.5mu#1\mkern-1.5mu}\mkern 1.5mu}

\begin{document}
\bstctlcite{IEEEexample:BSTcontrol}
    \title {Jump Plus AM-FM Mode Decomposition}
  \author{Mojtaba~Nazari, Anders Rosendal Korshøj,
      Naveed ur Rehman,~\IEEEmembership{Senior Member,~IEEE}
  \thanks{M. Nazari and N. Rehman are with the Department of Electrical and Computer Engineering, Aarhus University, 8200 Aarhus N, Denmark e-mails: m.nazari@ece.au.dk and naveed.rehman@ece.au.dk} 
  \thanks{Anders Rosendal Korshøj is with 1: Aarhus University Hospital, Dept. of Neurosurgery, Aarhus, Denmark, 2: Aarhus University, Dept. of Clinical Medicine, Aarhus, Denmark e-mail: anders.r.korshoj@clin.au.dk.}}%

\markboth{Submitted to IEEE Transactions on Signal Processing}%
{Mojtaba Nazari: Jump Plus AM-FM Mode Decomposition}

\maketitle

\begin{abstract}
A novel method for decomposing a nonstationary signal into amplitude- and frequency-modulated (AM-FM) oscillations and discontinuous (jump) components is proposed. Current nonstationary signal decomposition methods are designed to either obtain constituent AM-FM oscillatory modes or the discontinuous and residual components from the data, separately. Yet, many real-world signals of interest simultaneously exhibit both behaviors i.e., jumps and oscillations. Currently, no available method can extract jumps and AM-FM oscillatory components directly from the data. In our novel approach, we design and solve a variational optimization problem to accomplish this task. The optimization formulation includes a regularization term to minimize the bandwidth of all signal modes for effective oscillation modeling, and a prior for extracting the jump component. Our method addresses the limitations of conventional AM-FM signal decomposition methods in extracting jumps, as well as the limitations of existing jump extraction methods in decomposing multiscale oscillations. By employing an optimization framework that accounts for both multiscale oscillatory components and discontinuities, our methods show superior performance compared to existing decomposition techniques. We demonstrate the effectiveness of our approaches on synthetic, real-world, single-channel, and multivariate data, highlighting their utility in three specific applications: Earth's electric field signals, electrocardiograms (ECG), and electroencephalograms (EEG).
\end{abstract}

\begin{IEEEkeywords}
 Variational mode decomposition, multivariate data, empirical mode decomposition, biomedical applications, jump extraction 
\end{IEEEkeywords}

%
\IEEEpeerreviewmaketitle

\vspace{-3mm}
\section{Introduction}
\IEEEPARstart {S}ignal decomposition techniques are crucial for extracting and separating individual components from complex signals. Traditionally, nonstationary signal decomposition methods have focused on the extraction of amplitude-modulated and frequency-modulated (AM-FM) components \cite{Huang98,Dragomiretskiy14,daubechies2011synchrosqueezed}. The notable algorithms include Empirical Mode Decomposition (EMD) \cite{Huang98}, Variational Mode Decomposition (VMD) \cite{Dragomiretskiy14}, and Synchrosqueezed Transform (SST) \cite{daubechies2011synchrosqueezed}. These techniques decompose non-stationary signals into their constituent intrinsic mode functions (IMFs) and have found numerous cross-disciplinary applications \cite{pang2022recursive, akbari2022identification, gupta2015baseline}. Extensions for multivariate signals have also been developed \cite{Rehman10, Rehman19}.

\begin{figure}
    \centering
    \includegraphics[width=.5\textwidth]{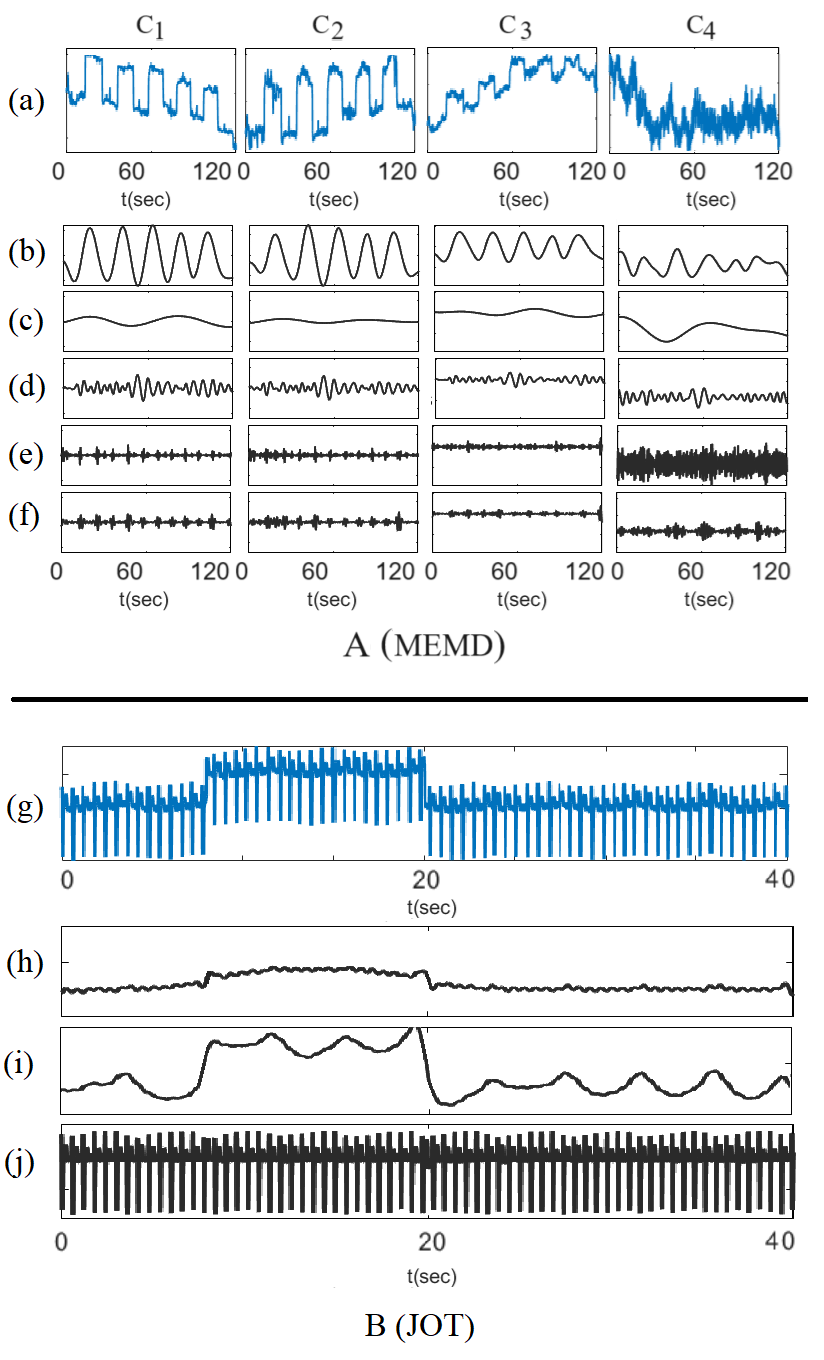}
    \caption{Two examples of real-world signals with jump components: multichannel electroencephalography (EEG) signal (a) and the single channel ECG signal (g). The decomposition of these two real-world signals into their AM-FM components using MEMD (A) and JOT (B). The MEMD wrongly models step-like jumps as the oscillatory components as illustrated in row (b). JOT also fails to decompose the AM-FM modes, as they are all mixed up, as shown in row (j).}
    \label{example}
\end{figure}

Despite their effectiveness, EMD and VMD primarily focus on the extraction of oscillatory components from the data. However, in many practical applications, data often displays a combination of oscillations, abrupt discontinuities (or jumps), and concurrent trends. Fig. \ref{example} (a) and (g), demonstrate two practical cases where jumps and oscillations are part of the signal and it is of interest to decompose such signals into their constituent components. In such signals, the discontinuities in the data significantly degrade the performance of traditional decomposition algorithms \cite{stallone2020new}. This is demonstrated in Fig. \ref{example} (b), where the output of the multivariate EMD (MEMD) algorithm mistakenly models a jump as an oscillation. That happens because, in the frequency domain, a sudden jump has a wide-band spectrum, affecting all frequency components. Consequently, filtering methods that use cutoff frequencies to distinguish jumps struggle to remove such artifacts effectively. These anomalies can affect all frequency components, making it difficult to separate them from the meaningful parts of the signal.

To address these challenges, a class of algorithms has been developed to handle jumps and noise more effectively \cite{storath2019smoothing, gholami2013balanced, kim2009ell_1, storath2014jump, selesnick2020non}. Particularly, \cite{storath2019smoothing} presents a method that estimates both discontinuities and a corresponding piecewise-smooth signal. Another study \cite{gholami2013balanced} combines Tikhonov and total variation regularizations to reconstruct piecewise-smooth signals by considering the signal as a sum of piecewise constant and smooth components. Researchers in \cite{kim2009ell_1} introduce the $\ell_1$ trend filtering method, which is a variant of Hodrick–Prescott filtering for estimating piecewise linear trends in time series data. Further advancements include the inverse Potts energy functionals \cite{storath2014jump}, which use dynamic programming and the alternating direction method of multipliers (ADMM) to recover jump-sparse and sparse signals from noisy, incomplete data. This technique overcomes the limitations associated with the $\ell_1$ norm by employing non-convex sparse regularizers (i.e., $\ell_0$ pseudonorm-based penalty), demonstrating superior performance in signal reconstruction compared to classical and recent approaches such as TV minimization \cite{needell2013stable}, orthogonal matching pursuit \cite{tropp2004greed}, and iterative hard thresholding \cite{blumensath2009iterative}.

The authors of \cite{cicone2022jot} proposed the Jumps, Oscillations, and Trends (JOT) model to decompose signals into three components. This model was inspired by the image decomposition work in \cite{meyer2001oscillating}, which proposed decomposing an image into its geometric part and its oscillatory texture or noisy part. The JOT method effectively separates signal components without artifacts, making it suitable for preprocessing. However, JOT has limitations in obtaining AM-FM oscillatory components since it considers all potential oscillations as a single component. Consequently, mid-low oscillatory components must be modeled either as high-frequency oscillations or as trends. Additionally, being a two-stage method, JOT has many input parameters that are challenging to tune. This tuning becomes even more complex when JOT is combined with other decomposition methods, each with their own set of parameters.

Here, we develop and present a novel method that jointly decomposes a nonstationary signal into jump and oscillatory (AM-FM) modes. Mathematically, the input signal is assumed to follow this model: 
\begin{equation}
\begin{aligned}
f(t)=\sum_{k=1}^K u_k(t) + v(t) + n(t),
\label{sigmodel}
\end{aligned}
\end{equation}

\noindent where $u_k(t)$ denotes the $K$ number of oscillatory or AM-FM components, $v(t)$ represents the jump component in the input data, and $n(t)$ refers to noise. The main goal of this work is to obtain $u_k(t)$ and $v(t)$, given the input signal $f(t)$. We achieve this by developing an optimization formulation that includes two different priors: one for obtaining the signal's AM-FM modes and another for its discontinuities. One penalty term enforces the AM-FM modes to be band-limited, while the other minimizes a function of the derivative of the jump component to precisely compute the amplitude of the piecewise-constant component. The proposed method is built on a single-stage algorithm rather than a two-stage process, offering a robust solution with fewer input parameters and a higher likelihood of convergence. The resulting optimization problem is solved using the ADMM approach. In the sequel, the proposed method is termed as Jump Plus AM-FM Mode Decomposition (JMD).

The rest of the paper is organized as follows: Section II reviews the related works and methods and their limitations in acquiring AM-FM modes and jumps. In section III we present the steps to build our formulation and solution to the problem, followed by several experiments and applications of our methods to both synthetic and real-life signals in section IV. Finally, discussions and concluding remarks are given in sections V and VI.

\section{Related Works and Limitations}
\subsection{AM-FM Mode Decomposition Methods}

Mode decomposition methods aim to decompose a signal into a finite number of amplitude- and frequency-modulated (AM-FM) modes. Let $f(t)$ denote an input signal, then the AM-FM decomposition methods obtain the modes $u_k(t)=a_k(t)$cos$(\phi_k(t))$, assuming the following model of the signal $f(t)$:
\begin{equation}
\begin{aligned}
f(t) = \sum_{k=1}^K u_k(t)
\label{vmdmodel}
\end{aligned}
\end{equation}

While several methods have been proposed to extract AM-FM modes from a nonstationary signal, the most widely used are EMD \cite{Huang98}, VMD \cite{Dragomiretskiy14}, and SST \cite{daubechies2011synchrosqueezed}. Invariably, all mode decomposition methods have been designed to process data containing smooth, oscillatory, and noise components only and their performance suffers significantly when discontinuities or jumps are part of the input signal. Specifically, the mode decomposition methods attempt to approximate jumps through oscillatory modes, resulting in spurious components (see row (b) in Fig. \ref{example}).

In their original formulations, AM-FM mode decomposition methods can only handle univariate or single-channel data sets. However, many real-world signals contain multiple channels exhibiting complex relationships e.g., the EEG signal shown in Fig. \ref{example} (a). To effectively process such data, several multivariate extensions of the AM-FM mode decomposition methods have emerged \cite{Rehman10,Rehman19}. Unlike their univariate counterparts, these methods can model cross-channel dependencies in multivariate data leading to improved performances in wide-ranging applications \cite{ur2010application,gupta2015baseline,lv2016multivariate}. However, similar to their univariate counterparts, the multivariate VMD (MVMD) and multivariate EMD (MEMD) are not able to effectively extract jump components from data. For example, Fig. \ref{example} (b) - (f) demonstrate the weak performance of MEMD in extracting the jump component from the multichannel signal (a), where the jump component (b) is inaccurately modeled as an oscillation.

\subsection{Jump Decomposition Methods}

Jump decomposition methods aim to accurately separate jump from signals \cite{storath2019smoothing, gholami2013balanced, kim2009ell_1, storath2014jump, selesnick2020non,needell2013stable,tropp2004greed,blumensath2009iterative}. These methods are crucial for identifying and removing anomalies, such as jumps or spikes, which can obscure the meaningful analysis of the underlying signal trends and frequencies. Notably, Potts focuses on partitioning the signal into segments with distinct levels, effectively capturing jumps \cite{storath2014jump}. This approach uses an $\ell_0$ pseudonorm-based penalty, making it robust against noise and capable of preserving sharp transitions. More recently, the two-stage approach JOT has been proposed that aims to decompose signals into three distinct components: Jumps ($v^*$), Oscillations ($n^*$), and Trends ($w^*$) \cite{cicone2022jot} as follows: 
\begin{equation}
\begin{aligned}
f(t) = v^*(t) + n^*(t) + w^*(t).
\label{JOTmodel}
\end{aligned}
\end{equation}

The method first identifies these components by solving a non-convex minimization problem to capture jumps and oscillations, then refines these components by redistributing trends and oscillatory residuals. Although JOT and Potts are effective in separating signal components without artifacts, they struggle with extracting AM-FM oscillatory modes due to their signal models, which treat all oscillations as a single component. For instance, Fig. \ref{example} (h) - (j) demonstrates the limitation of JOT in extracting the AM-FM modes from a single-channel ECG signal with jump artifacts (g). In addition to the mode mixing problem, all oscillatory components are inaccurately modeled as a single oscillation (as illustrated in row (j)).

\section{Jump Plus AM-FM Mode Decomposition}

In this section, we present our proposed variational JMD model to extract AM-FM modes $u_k(t)$ and jump component $v(t)$, from an input signal $f(t)$, in accordance with the signal model given in \eqref{sigmodel}. Our approach aims to achieve the following two goals: i) extracting the jumps by incorporating a sparsity-inducing regularizer, and ii) decomposing the signal into a finite number of oscillatory components (sub-signals or modes) by minimizing a term related to the bandwidth of the modes. This combination leads to a minimization problem for effectively separating the discontinuity features from oscillating parts of the original signal. 

In this framework, the assumption is that each oscillatory mode is predominantly localized around a central frequency \cite{Dragomiretskiy14} while the discontinuities are wide-band components on the spectrum. Considering these two opposite features, incorporating two different terms in our formulation is required. We will discuss this further in the following subsections.

\subsection{Oscillatory Components}
In order to obtain the AM-FM oscillatory components from the data, we use the following optimization formulation of the VMD algorithm that aims to minimize the bandwidth of the AM-FM components:
\begin{equation}
\small
\begin{aligned}
J_1= \sum_k\Big\Vert\partial_t\Big[\tilde{u}_{k+} (t) e^{-j\omega_{k} t}\Big]\Big\Vert^2_2,
\label{vmdcost}
\end{aligned}
\end{equation}

\noindent where the symbol $\partial_t$ represents the partial derivative with respect to time, $u_{k+}(t) = u(t) + j \mathcal{H} u(t) = a(t) e^{j \phi(t)}$ (and $\tilde{u}_+(\omega) = (1 + \text{sgn}(\omega)) \tilde{u}(\omega)$) refers to the analytic signal in which $\mathcal{H}$ denotes the Hilbert transform, and ${\omega_k}$ is the center frequency of each mode. In essence, the squared $l^2$-norm term estimates the bandwidths of the $k$-th oscillatory modes of the signal. The minimization of the bandwidth of a mode is tantamount to approximating it to a sinusoidal signal, which by definition has the minimum possible bandwidth.

\subsection{Jump Component}

To extract the jump component of a signal, we use a reparameterized and rescaled minimax concave (MC) penalty term, originally proposed in \cite{zhang2010nearly}. This penalty term penalizes the derivative of the jump component to promote a piecewise-constant component. It aims to accurately preserve the amplitude of the piecewise-constant signal component while improving the representation of sharp discontinuities in the jump component: 
\begin{equation}
\small
\begin{aligned}
J_2= \int_{0}^\infty \phi (\Vert \partial_t v(t) \Vert_2; b) \quad dt,
\label{jumpcost}
\end{aligned}
\end{equation}
\noindent where $\partial_t v(t) := v'(t)$ represents the first derivative of the jump component. The non-convex sparsity-promoting penalty function $\phi(\cdot; b) : [0, +\infty) \to [0, 1]$ is defined as a piecewise-quadratic function \cite{cicone2022jot}, as follows:
\begin{align}
\centering
\phi (x; b) &= \begin{cases}
-\frac{b}{2}x^2 + \sqrt{2b} x & x \in [0,\frac{2}{b}), \\
1 & x \in [\frac{2}{b}, +\infty).
\end{cases} &\quad
\label{phifunc}
\end{align}

In the above function, parameter $b$ changes the degree of non-convexity, such that for $b \xrightarrow{} 0$, the MC penalty is defined as $\phi (x; b) = |x|$ and $\phi (\cdot; b)$ tends to $l_0$-pseudonorm for $b \xrightarrow{} \infty$. 
Utilizing this term is motivated by an image decomposition study \cite{meyer2001oscillating}, which proposed decomposing the image into three parts: geometric parts, and oscillatory texture or noisy part by means of TV denoising \cite{rudin1992nonlinear}. This formulation has also been used successfully to extract jump components from input data \cite{cicone2022jot,selesnick2020non}. 

\subsection{Optimization Formulation}

In our proposed solution to obtain $v(t)$ and $u_k(t)$ using input data $f(t)$ in \eqref{sigmodel}, we combine \eqref{vmdcost} and \eqref{jumpcost} within a single optimization formulation as follows:
\begin{equation}
\begin{split}
\begin{aligned}
\underset{\{{\tilde{u}}_k\},\{\omega_{k}\},\{v\}}{\text{minimize}}
\footnotesize\Big\{ \alpha{J}_1 + \beta{J}_2\Big\},
\label{opt1} 
\end{aligned}
\end{split}
\end{equation}
\begin{equation}
\small
 \centering 
s.t. \quad
v + \sum_k u_{k}= f.
\label{cosnt} 
\end{equation}
Next, we perform a variable splitting technique by defining an auxiliary variable $x:= \partial_t v$ to solve the problem of non-differentiability of the term $\phi ( \cdot ; b)$. Then we can write our problem in this form: 
\begin{equation}
\begin{split}
\begin{aligned}
\underset{\{{{u}}_k\},\{\omega_{k}\},\{v\}}{\text{minimize}}
\footnotesize\Big\{ \alpha\sum_k\Big\Vert\partial_t\Big[{u}_{k+} (t) e^{-j\omega_{k} t}\Big]\Big\Vert^2_2 \\ 
+ \beta \int_{0}^\infty \phi (\Vert \partial_t v(t) \Vert_2; b) \hspace{1mm} dt \Big\},
\label{opt2} 
\end{aligned}
\end{split}
\end{equation}
\begin{equation}
\small
 \centering 
s.t.\begin{cases}
v + \sum_k u_{k}= f,  
\\
x= \partial_t v
\label{jointcosnt} 
\end{cases}
\end{equation}
The above optimization formulation is constrained and cumbersome to solve. We convert it into unconstrained formulation using Lagrangian multipliers, resulting in the following augmented Lagrangian function:
\begin{equation}
\small
\begin{aligned}
\mathcal{L}\{{{u}}_k,\omega_{k},v,x,\rho, \lambda\} = \alpha\sum_k\Big\Vert\partial_t\Big[{u}_{k+} (t) e^{-j\omega_{k} t}\Big]\Big\Vert^2_2  
+ \Big\Vert f(t) - \\
\Big(v(t) + \sum_k u_k(t)\Big)\Big\Vert^2_2 + \Big\langle \lambda(t), f(t) - \Big(v(t) + \sum_k u_k(t)\Big)\Big\rangle 
\\ + \beta \int_{0}^\infty \phi (\Vert x(t) \Vert_2; b) dt - \Big\langle \rho(t), x(t) - \partial_t v(t)\Big\rangle + \frac{\gamma}{2}\Big\Vert x(t) - \\ 
\partial_t v(t)\Big\Vert^2_2,
\label{lagran}
\end{aligned}
\end{equation}

\noindent where $\gamma$ is the penalty scalar parameter, and $\lambda(t)$ and $\rho(t)$ are the vectors of Lagrangian multipliers associated with constraints  $\sum_k u_{k}+v= f$ and $x= \partial_t v$, respectively.

\subsection{Optimization Solution}
The above formulation is solved using the alternate direction method of multipliers (ADMM). To accomplish that, we divide the augmented Lagrangian formulation into several sub-optimization problems and solve those separately and iteratively.

\subsubsection{Minimization w.r.t $u_k$, $\omega_k$, and $\lambda$}
The sub-optimization problem to solve for $u_k$, $\omega_k$, and $\lambda$ can be obtained from \eqref{lagran} by including only those terms that depend on them:
\begin{equation}
\small
\begin{aligned}
\mathcal{L}\{{{u}}_k,\omega_{k}, \lambda\} = \alpha\sum_k\Big\Vert\partial_t\Big[{u}_{k+} (t) e^{-j\omega_{k} t}\Big]\Big\Vert^2_2  
+ \Big\Vert f(t) - \\
\Big(v(t) + \sum_k u_k(t)\Big) +\frac{\lambda(t)}{2} \Big\Vert^2_2
\label{lagran2}.
\end{aligned}
\end{equation}

As proposed in \cite{Dragomiretskiy14}, by solving \eqref{lagran2} in the Fourier domain, taking the partial derivative, and some algebraic manipulations, the associated update equations will be obtained as follows:
\begin{equation}
\small
\begin{aligned}
\hat{u}_{k}^{(i+1)}(\omega) = \frac{\hat{f}(\omega) - \sum_{l\neq k} u_l^{(i)}(\omega) - \hat{v}^{(i)}(\omega)+\frac{\hat{\lambda}^{(i)}(\omega)}{2}}{1+2\alpha(\omega-\omega_{k}^{(i)})^2},
\end{aligned}
\label{uk}
\end{equation}

 \begin{equation}
 \small
\omega_{k}^{i+1}=\frac{\mathlarger\int^{\infty}_0 \omega \Big|\hat{u}_{k}^{(i+1)}(\omega)\Big|^2 d\omega}{\mathlarger\int^{\infty}_0 \Big|\hat{u}_{k}^{(i+1)}(\omega)\Big|^2 d\omega},
\label{omega}
\end{equation}

\begin{equation}
\small
\begin{aligned}
\lambda^{(i+1)}=\lambda^{(i)}(\omega)+\tau_1 \Bigg( {\hat{f}}(\omega)-\Bigg[{\hat{v}}^{(i)}(\omega)+\sum_{k} u_k^{(i+1)}(\omega)\Big)\Bigg]\Bigg).\hspace{5mm}
\label{lambda}
\end{aligned}
\end{equation}

\subsubsection{Minimization w.r.t $v$, $x$, and $\rho$}
The suboptimization problem to solve for $v$, $x$, and $\rho$ can be obtained from \eqref{lagran} and is given as follows:
\begin{equation}
\small
\begin{aligned}
\mathcal{L}\{v,x,\rho\} = \beta \int_{0}^\infty \phi (\Vert x(t) \Vert_2; b) dt - \Big\langle \rho(t), x(t) - \partial_t v(t)\Big\rangle \\
+ \frac{\gamma}{2}\Big\Vert x(t) - \partial_t v(t)\Big\Vert^2_2 + \Big\Vert f(t) - 
\Big(v(t) + \sum_k u_k(t)\Big)\Big\Vert^2_2 \\ +\Big\langle \lambda(t), f(t) - \Big(v(t) + \sum_k u_k(t)\Big)\Big\rangle.
\label{lagran3_0}
\end{aligned}
\end{equation}

This sub-problem is cumbersome to solve directly in the continuous domain owing to the first term involving the infinite integral. Fortunately, the term involving $\phi$ can be converted to a discrete form and has been shown to be strongly convex under necessary conditions in \cite{huska2019convex}. In the following, we describe the steps to convert \eqref{lagran3_0} in the discrete form and solve it to obtain the update equations for $v$, $x$, and $\rho$. 

Assume that we deal with a finite signal\footnote[1]{Although we make this assumption to simplify the sub-function including $\phi ( \cdot ; b)$ and proceed in the discrete domain; practically this is also numerically established for \eqref{lagran2} as explained in \cite{Dragomiretskiy14}.}. Then, we can rewrite the optimization function \eqref{lagran} to minimize w.r.t $v$, $x$ and $\rho$ as follows: 
\begin{equation}
\small
\begin{aligned}
\mathcal{L}\{v,x,\rho\}= \beta \sum_{j=1}^{N} \phi (\Vert x_j \Vert_2; b) - \Big\langle \rho, x - Dv\Big\rangle + \frac{\gamma}{2}\Big\Vert x -Dv\Big\Vert^2_2 \\
+ \Big\Vert f - \Big(v + \sum_k u_k\Big)\Big\Vert^2_2 + \Big\langle \lambda, f - \Big(v + \sum_k u_k\Big)\Big\rangle,
\label{lagran3}
\end{aligned}
\end{equation}

\noindent where $D$ is defined as the first-order derivative matrix as follows:
\begin{equation}
\small
\begin{aligned}
D = \begin{bmatrix}
-1 & 1 \\
 & -1 & 1 \\
 & & \ddots & \ddots \\
 & & & -1 & 1
\end{bmatrix} \in \mathbb{R}^{(N-1) \times N}.
\label{derivmat}
\end{aligned}
\end{equation}

Next, by solving \eqref{lagran3}, taking the partial derivative with respect to $v$, and zero-initializing the vectors $x^{(0)}$ and $\rho^{(0)}$, we can obtain the $i$-th iteration of $v$ as follows:
\begin{equation}
\small
\begin{aligned}
v^{(i+1)} = (\gamma D^TD + 2I)^{-1} \Big(D^T\rho^{(i)} + \gamma D^T x^{(i)} + \\2(f-\sum_k u_k^{(i+1)}) - \lambda^{(i+1)}\Big).
\label{vjump}
\end{aligned}
\end{equation}
Moreover, we can write the subproblem for obtaining $x$ by omitting the constant terms as:
\begin{equation}
\small
\begin{aligned}
x^{(i+1)} = \arg\min_{x \in \mathbb{R}^N} \Big\{ \beta \sum_{n=1}^N \phi(|x_j|; b) - \langle \rho, x \rangle + \frac{\gamma}{2} \lVert x - Dv \rVert_2^2 \Big\}.
\label{xlagran}
\end{aligned}
\end{equation}

The above optimization problem, when expressed in terms of its individual components, can be seen as comprising $N$ separate one-dimensional problems:
\begin{equation}
\small
\begin{aligned}
x_j^{(i+1)} = \arg\min_{x \in \mathbb{R}} \Big\{  \mu \phi(|x|; b) + \frac{1}{2} \lVert x - (Dv^{(i)})_j + \frac{\rho^{(i)}_j}{\gamma} \rVert_2^2 \Big\},\\ j = 1, \ldots, N,
\label{xder}
\end{aligned}
\end{equation}

\noindent where $\mu = \frac{\beta}{\gamma}$. As demonstrated in \cite{huska2019convex}, the problems in \eqref{xder} are strongly convex if and only if sufficient conditions for strong convexity are met:
\begin{equation}
\small
\begin{aligned}
b < \frac{1}{\mu} \Rightarrow \gamma > b\beta \Rightarrow \gamma = \tau_2 b\beta, \text{ for } \tau_2 \in \mathbb{R}, \tau_2 > 1.
\label{convcon}
\end{aligned}
\end{equation}

Given assumption \eqref{convcon}, closed-form solutions for the problems stated in \eqref{xder} can be derived as:
\begin{equation}
\small
\begin{aligned}
x_j^{(i+1)} =\min \left( \max \left( \frac{1}{1 - \mu b} - \frac{\frac{\mu\sqrt{2b}}{1-b\mu}}{\lvert h_j \rvert}, 0 \right), 1 \right)  h_j,
\label{xupdate}
\end{aligned}
\end{equation}

\noindent where $h_j= (Dv^{(i)})_j + \frac{\rho^{(i)}_j}{\gamma}$. Finally, the update for $\rho$ will be governed by the following equation: 
\begin{equation}
\small
\begin{aligned}
\rho^{(i+1)} = \rho^{(i)} - \gamma \left( x^{(i+1)} - Dv^{(i+1)} \right)
\label{rho}
\end{aligned}
\end{equation}

With all relevant update equations at our disposal, we list the steps for obtaining the AM-FM modes $u_k(t)$ and jump component $v(t)$, given an input signal $f(t)$, in Algorithm \ref{JMD}.

\begin{algorithm}
    \caption{\bf JMD}
    \label{JMD}
    \small
    \textbf{Input and user-defined parameters:} $f, K, \alpha, \beta, \overline{b}$ \\
    \vspace{1mm}
    \textbf{Initialize:} $\epsilon \gets 10^{-7}, \hat{u}_{k}^{(1)}, \omega_k^{(1)}, x^{(1)}, \lambda^{(1)}, \rho^{(1)} \gets 0$ \\
    \vspace{0mm}
    \textbf{Generate:} Discrete operator $D$ in \eqref{derivmat}
    \begin{algorithmic}[1]
        \State $b = 2/{\overline{b}^2}, \quad \gamma = \tau_2 b \beta$
        \vspace{1mm}
        \Repeat
            \State $i \gets i+1$
            \vspace{1mm}
            \For{$k=1:K$}
                \State \textit{Update mode $\hat{u}_{k}$:}
                \vspace{1mm}
                \State $\hat{u}_{k}^{(i+1)}(\omega) \gets \text{Using Equation \eqref{uk}}$
                \vspace{1mm}
                \State \textit{Update $\omega_{k}$:}
                \vspace{1mm}
                \State $\omega_{k}^{(i+1)}\gets\text{Using Equation \eqref{omega}}$
                \vspace{1mm}
            \EndFor
            \vspace{1mm}
            \State $\sum_k u_k\xleftarrow{IFT}\sum_k \hat{u}_k(\omega)$ \textit{(to compute $v$)}
            \vspace{1mm}
            \State \textit{Update $\lambda, v, x,$ and $\rho$:}
            \vspace{1mm}
            \State $\lambda^{(i+1)}(\omega)\gets\text{Using Equation \eqref{lambda}}$
            \vspace{1mm}
            \State $v^{(i+1)}\gets\text{Using Equation \eqref{vjump}}$
            \vspace{1mm}
            \State $x^{(i+1)}\gets\text{Using Equation \eqref{xupdate}}$
            \vspace{1mm}
            \State $\rho^{(i+1)}\gets\text{Using Equation \eqref{rho}}$
            \vspace{1mm}
            \State $r^{(i+1)} = \sum_k {u}_{k}^{(i+1)} + {v}^{(i+1)}$
            \vspace{1mm}
            \State $\hat{v}^{(i+1)}(\omega)\xleftarrow{FT} v^{(i+1)}$ \textit{(to update $\hat{u}_k(\omega)$)}
            \vspace{1mm}
        \Until{Convergence: $ \Vert r^{(i+1)}-r^{(i)}\Vert_2^2/\Vert r^{(i)}\Vert_2^2<\epsilon$}
    \end{algorithmic}
\end{algorithm}

\subsection{Multivariate Extension}

Here we present the multivariate extension of the JMD method. We start with the multivariate extension of the jump and AM-FM mode decomposition model given in \eqref{sigmodel}. The model is given as follows:
\begin{equation}
\bold {f}(t) = \sum_{k=1}^{K} \mathbf{u}_{k}(t) + \bold v(t) + \bold n(t),
\label{sigmodelmulti}
\end{equation}
\noindent where $\bold f(t)$ denotes a multivariate signal consisting of C number of data channels $\bold f(t)=[f_1(t),...f_C(t)]$. The signal can also be represented by a matrix $\bold F$ in the discrete $\mathbf{F} \in \mathbb{R}^{C \times N}$. Further, $\bold u_k(t)$ are the set of AM-FM decomposed modes, all of which are multivariate with $C$ number of data channels. Finally, $\bold v(t)$ and $\bold n(t)$ refer to the multivariate jump component in the data and noise, respectively.

The goal of the multivariate extension of the JMD method is to extract the jump components $\bold v(t)$ in all channels and the AM-FM modes $\bold u_k(t)$ from input signal $\bold f(t)$. Unlike the univariate algorithm, the AM-FM modes will be aligned in frequency in the multivariate algorithm, enabling the direct comparison between the modes from different channels \cite{Rehman19}. To achieve this, we utilize the cost function of the MVMD method in our problem such that the bandwidth of $\bold u_k(t)$ can be determined by calculating the $\ell_2$ norm of the gradient of the harmonically shifted $\bold u_k^+(t)$ \cite{Rehman19}:
\begin{equation}
\begin{aligned}
J_1=\alpha\mathlarger{\sum_{k}\sum_c}\Bigg\Vert\partial_t\Big[{u}_{k,c}^+(t) e^{-j\omega_{k}t}\Big]\Bigg\Vert^2_2,
\label{mvmd_opt}
\end{aligned}
\end{equation}

\noindent where $c=1,2,\ldots, C$ denote number of channels. Said that, the term for obtaining jumps will not be affected; thus, it remains almost the same:

\begin{equation}
\begin{aligned}
J_2= \beta \int_{0}^\infty \phi (\Vert \partial_t v_c(t) \Vert_2; b) \hspace{1mm} dt.
\label{jumpmulti}
\end{aligned}
\end{equation}

Utilizing the same procedure of our univariate JMD optimization formulation i.e. in \eqref{opt2} and \eqref{jointcosnt}, we can formulate the problem as follows:
\begin{equation}
\begin{split}
\begin{aligned}
\underset{\{u_{k,c}\},\{\omega_{k}\},\{v_{c}\}}{\text{minimize}}
\footnotesize\Big\{ \alpha\sum_k \sum_c\Big\Vert\partial_t\Big[{u}_{k,c}^+ (t) e^{-j\omega_{k} t}\Big]\Big\Vert^2_2 \\ 
+ \beta \int_{0}^\infty \phi (\Vert \partial_t v_c(t) \Vert_2; b) \hspace{1mm} dt \Big\}
\label{mJMD} 
\end{aligned}
\end{split}
\end{equation}

\begin{equation}
\small
 \centering 
s.t.\begin{cases}
v_c + \sum_k u_{k,c}= f_c  
\\
x_c= \partial_t v_c 
\end{cases}.
\label{mjointcosnt} 
\end{equation}

Similarly, we convert the above optimization formulation into an unconstrained formulation using Lagrangian multipliers, resulting in the following augmented Lagrangian function:
\begin{equation}
\small
\begin{aligned}
\mathcal{L}\{{{u}}_{k,c},\omega_{k},v_c,x_c,\rho_c, \lambda_c\} = \alpha\sum_k \sum_c\Big\Vert\partial_t\Big[{u}_{k,c}^+ (t) e^{-j\omega_{k} t}\Big]\Big\Vert^2_2 \\ 
+ \sum_c \Big\Vert f_c(t) - 
\Big(v_c(t) + \sum_k u_{k,c}(t)\Big)\Big\Vert^2_2 + \sum_c\Big\langle \lambda_c(t), f_c(t) - \\ \Big(v_c(t) + \sum_k u_{k,c}(t)\Big)\Big\rangle + \beta \int_{0}^\infty \phi (\Vert x_c(t) \Vert_2; b) dt \\- \Big\langle \rho_c(t), x_c(t) - \partial_t v_c(t)\Big\rangle + \frac{\gamma}{2}\Big\Vert x_c(t) -
\partial_t v_c(t)\Big\Vert^2_2.
\label{lagranmulti}
\end{aligned}
\end{equation}

The above Lagrangian function can be similarly solved using ADMM as our presented procedure for the univariate JMD method in previous subsections (i.e., in equations \eqref{lagran2} to \eqref{rho}). Therefore, the relevant update equations for $u_{k,c}$, $\omega_k$, $\lambda_c$, $v_c$, $x_c$, and $\rho_c$ will be given as follows:
\begin{equation}
\small
\begin{aligned}
\hat{u}_{k,c}^{(i+1)}(\omega) = \frac{\hat{f_c}(\omega) - \sum_{l\neq k} u_{l,c}^{(i)}(\omega) - \hat{v_c}^{(i)}(\omega)+\frac{\hat{\lambda_c}^{(i)}(\omega)}{2}}{1+2\alpha(\omega-\omega_{k}^{(i)})^2},
\end{aligned}
\label{muk}
\end{equation}

 \begin{equation}
 \small
\omega_{k}^{i+1}=\frac{\mathlarger{\sum_c}\mathlarger\int^{\infty}_0 \omega \Big|\hat{u}_{k,c}^{(i+1)}(\omega)\Big|^2 d\omega}{\mathlarger{\sum_c}\mathlarger\int^{\infty}_0 \Big|\hat{u}_{k,c}^{(i+1)}(\omega)\Big|^2 d\omega},
\label{momega}
\end{equation}

\begin{equation}
\small
\begin{aligned}
\lambda_c^{(i+1)}=\lambda_c^{(i)}(\omega)+\tau_1 \Bigg( {\hat{f_c}}(\omega)-\Bigg[{\hat{v_c}}^{(i)}(\omega)+\sum_{k} u_{k,c}^{(i+1)}(\omega)\Big)\Bigg]\Bigg),\hspace{5mm}
\label{mlambda}
\end{aligned}
\end{equation}
\begin{equation}
\small
\begin{aligned}
v_c^{(i+1)} = (\gamma D^TD + 2I)^{-1} \Big(D^T\rho_c^{(i)} + \gamma D^T x_c^{(i)} + \\2(f_c-\sum_{k} u_{k,c}^{(i+1)}) - \lambda_c^{(i+1)}\Big),
\label{mvjump}
\end{aligned}
\end{equation}
\begin{equation}
\small
\begin{aligned}
x_{c,(j)}^{(i+1)} =\min \left( \max \left( \frac{1}{1 - \mu b} - \frac{\frac{\mu\sqrt{2b}}{1-b\mu}}{\lvert h_{c,(j)} \rvert}, 0 \right), 1 \right)  h_{c,(j)},
\label{mxupdate}
\end{aligned}
\end{equation}

\noindent and
\begin{equation}
\small
\begin{aligned}
\rho_{c}^{(i+1)} = \rho_c^{(i)} - \gamma \left( x_c^{(i+1)} - Dv_c^{(i+1)} \right).
\label{mrho}
\end{aligned}
\end{equation}

Having all these update equations, the multivariate JMD algorithm can be summarized in Algorithm \ref{MJMD}.

\begin{algorithm}
    \caption{\bf MJMD}
    \label{MJMD}
    \small
    \textbf{Input and user-defined parameters:} $f, K, \alpha, \beta, \overline{b}$ \\
    \vspace{1mm}
    \textbf{Initialize:} $\epsilon \gets 10^{-7}, \hat{u}_{k,c}^{(1)}, \omega_k^{(1)}, x_c^{(1)}, \lambda_c^{(1)}, \rho_c^{(1)} \gets 0$ \\
    \vspace{1mm}
    \textbf{Generate:} Discrete operator $D$ in \eqref{derivmat}
    \begin{algorithmic}[1]
        \State $b = 2/{\overline{b}^2}, \quad \gamma = \tau_2 b \beta$
        \vspace{1mm}
        \Repeat
        \vspace{1mm}
            \State $i \gets i+1$
            \vspace{1mm}
            \For{$k=1:K$}
            \vspace{1mm}
                \For{$c=1:C$}
                \vspace{1mm}
                    \State \textit{Update mode $\hat{u}_{k,c}$:}
                    \vspace{1mm}
                    \State $\hat{u}_{k,c}^{(i+1)}(\omega) \gets \text{Using Equation \eqref{muk}}$
                    \vspace{1mm}
                \EndFor
                \vspace{1mm}
                \State \textit{Update $\omega_{k}$:}
                \vspace{1mm}
                \State $\omega_{k}^{(i+1)}\gets\text{Using Equation \eqref{momega}}$
                \vspace{1mm}
            \EndFor
            \vspace{1mm}
            \For{$k=1:K$}
            \vspace{1mm}
                \For{$c=1:C$} 
                \vspace{1mm}
                    \State $u_{k,c}\xleftarrow{IFT} \hat{u}_{k,c}(\omega)$ \textit{(to compute $v_c$)}
                    \vspace{1mm}
                \EndFor
                \vspace{1mm}
            \EndFor
            \vspace{1mm}
            \State \textit{Update $\lambda_c, v_c, x_c,$ and $\rho_c$:}
            \vspace{1mm}
            \For{$c=1:C$}
            \vspace{1mm}
                \State $\lambda_c^{(i+1)}(\omega)\gets\text{Using Equation \eqref{mlambda}}$
                \vspace{1mm}
                \State $v_c^{(i+1)}\gets\text{Using Equation \eqref{mvjump}}$
                \vspace{1mm}
                \State $x_c^{(i+1)}\gets\text{Using Equation \eqref{mxupdate}}$
                \vspace{1mm}
                \State $\rho_c^{(i+1)}\gets\text{Using Equation \eqref{mrho}}$
                \vspace{1mm}
                \State $\hat{v}_c^{(i+1)}(\omega)\xleftarrow{FT} v_c^{(i+1)}$ \textit{(to update $\hat{u}_{k,c}(\omega)$)}
                \vspace{1mm}
                \EndFor
                \vspace{1mm}
                    \For{$c=1:C$} 
                    \vspace{1mm}
                        \State $r_c^{(i+1)} = \sum_k{u}_{k,c}^{(i+1)} + {v_c}^{(i+1)}$
                        \vspace{1mm}
                    \EndFor
            \vspace{1mm}
        \Until{Convergence: $ \sum_c \Vert r_c^{(i+1)}-r_c^{(i)}\Vert_2^2/\Vert r_c^{(i)}\Vert_2^2<\epsilon$}
    \end{algorithmic}
\end{algorithm}

\subsection{Input Parameters}
Input parameters play a key role in the convergence of our presented methods. The parameters $K$ and $\alpha$ must be precisely determined in advance. A large $K$ can lead to duplicate modes, whereas a small $K$ causes mode mixing problems. Selecting a proper $K$ must be done by analyzing the signal in hand \textit{a priori}, and adaptive strategies have also been used to tune $K$ \cite{zhang2018parameter,li2020optimized,lian2018adaptive}. Similarly, a high value of $\alpha$ can result in incorrect noisy modes or cause convergence issues, while a low value of $\alpha$ can lead to mode mixing problems. Essentially, a high $\alpha$ value may be suitable when dealing with a dense spectrum with closely spaced frequency components. We found that setting the parameter $\alpha$ in the range of $10^3 - 10^5$ leads to satisfactory performance in most experiments. Additionally, different center frequency initialization strategies can cause methods like our presented JMD methods to diverge from the actual center frequencies in the spectrum \cite{Dragomiretskiy14}.
The parameter $\tau_1$ (in \eqref{lambda}) balances faithfully reconstructing and denoising the input signal. Opting for a smaller $\tau_1$ value may be advisable when addressing significant noise levels in the input signals. This parameter is set in the range of $0 - 1$.
Parameter $\beta$ is a regularization parameter that affects the extracted jump component. This parameter should be set approximately to the inverse of the number of expected jumps in the signal i.e., $1/\text{(number of expected jumps)}$. This ensures that the regularization term properly balances the contribution of the jump component without over-penalizing or under-penalizing it. Fine-tuning, however, is then needed to achieve the best performance for the specific applications. Parameter $\tau_2 > 1$ is typically set in the range of 1.1 to 50, with a default value of 10, for initializing $\gamma$. Parameter $\overline{b}$ refers to the expected minimal jump height, with a default value of 0.3. It indicates the minimal height above which every jump $\left| (Dv)_j \right|$ greater than $\overline{b}$ is equally penalized. The user should set $\overline{b}$ based on the expected minimal jump height observed in the input data. If every non-zero derivative $\left| (Dv)_j \right| > 0$ satisfies $\left| (Dv)_j \right| > \overline{b}$, the value of the $v$-regularization term will be equal to the number of non-zero derivatives (jumps) in $v$.

\section{Experiments and Results}

We employ the proposed algorithm on a series of synthetic and real-world signals to evaluate the effectiveness of our approach. Time is discretized into $t\in [0,1]$ in all simulations. We also compare the components obtained from the JMD and MJMD methods with the results obtained from the VMD \cite{Dragomiretskiy14}, MVMD \cite{Rehman19}, and the JOT \cite{cicone2022jot} models. Note that in this section, the amplitude of components (y axis) is scaled between $0 - 1$ in all figures for a better illustration of the results. Additionally, the presented methods are evaluated by setting the parameter $\tau_1=0$ as we assume that the input signal is corrupted by noise. For the comparative methods, we set the parameter values to ensure a fair comparison with the proposed methods. For instance, in VMD and MVMD, we use the same values for $K$ as those used in the proposed (M)JMD method. In other cases, we use the parameter values for the comparative methods that produce the best results.

\begin{figure}
\centering
\includegraphics[width=0.5\textwidth]{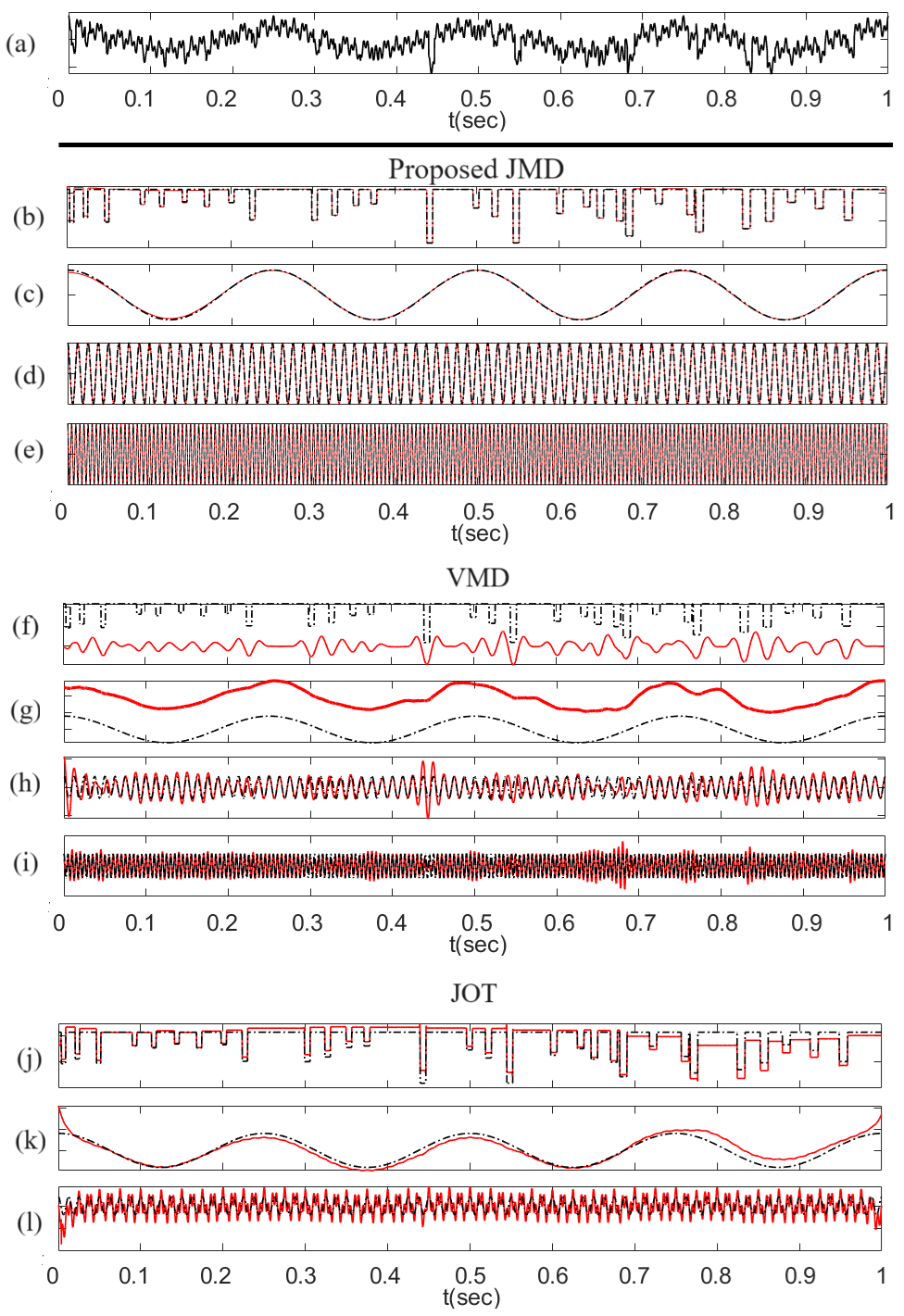}
\caption{Synthetic signal (a), and decomposed components obtained from the JMD method (rows (b) - (e)), VMD (rows (f) - (i)), and JOT (rows (j) - (l)). In all plots, the red line refers to the extracted components and the black dashed line indicates the original components of the signal.}
\label{Synth}
\end{figure}

\subsection{Synthetic Signal}
\subsubsection{Example 1 (Univariate Signal)}
To test the presented univariate model we start with a simple single-channel signal $f_1(t)$ that is composed of three 4Hz, 80Hz, and 200 Hz oscillatory components, one jump $v(t)$, and additive white Gaussian noise (AWGN) $\eta$ with zero mean and variances of $\sigma=0.1$ as follows:
\begin{equation}
\small
\begin{aligned}
f_1(t)&=\cos(2\pi 4 t) + \cos(2\pi 80 t) + \cos(2\pi 200 t) + v(t) + \eta.
\end{aligned} 
\end{equation}

The input parameters for this experiment were set to $K=3$, $\alpha=5000$, $\beta=0.03$, $\overbar{b}=0.45$, and $\tau_2=50$. 

Figure \ref{Synth} visualizes the extracted components from the proposed JMD method including three oscillatory modes and one jump. It is evident that our method was able to extract all components precisely. We also show the results of the comparative methods in this figure. For instance, VMD clearly fails to extract jumps as its formulation has not been designed to capture jumps. As the jump component has a broad spectrum, a broad spectral overlap is expected between the jump and the other three components. Due to this spectral overlap, VMD struggles to capture the pure modes as it performs like a Wiener filtering of the residual on the spectrum \cite{Dragomiretskiy14}. Although the JOT method is able to capture jumps, it cannot obtain oscillatory modes accurately. Especially, the 80Hz and 200Hz extracted components got mixed up in one component illustrated in Fig. \ref{Synth}- row (l). Because JOT decomposes the signal into three components including jump, oscillation, and trend, it only provides three components even in cases where there are more components than three.

\subsubsection{Example 2 (Multivariate Signal)}
To assess the performance of the proposed multivariate extension of the JMD method, we performed a series of experiments and simulations on various multivariate signals. The comprehensive results of these analyses are detailed in this section. 
We will focus on the method's ability to align common frequency scales across multiple data channels, which is crucial for various scientific and engineering applications involving multivariate data \cite{rehman2009bivariate,hao2017joint,gupta2015baseline}.
Here we showcase the mode-alignment property of MJMD using synthetic test signals composed of multiple oscillatory components across different channels and AWGN. We also compare these results with those obtained from MVMD and JOT. In the case of JOT, the results are obtained by applying this method to each channel separately. The synthetic multivariate data includes 3 channels (i.e., $C_1, C_2$, and $C_3$) as follows:
\begin{equation}
\small
\begin{aligned}
C_1(t)&=10\cos(2\pi 1 t) + 2\cos(2\pi 24 t) + v(t) + \eta
\\
C_2(t)&=4\cos(2\pi 1 t) + 2\cos(2\pi 24 t) + 2\cos(2\pi 48 t) \\
& + 2\cos(2\pi 128 t) + \eta
\\
C_3(t)&=10\cos(2\pi 1 t) + 2\cos(2\pi 48 t) + 2\cos(2\pi 128 t) + v(t) + \eta
\end{aligned} 
\end{equation}

The input parameters for this experiment were set to $K=3$, $\alpha=5000$, $\beta=0.05$, $\overbar{b}=0.45$, and $\tau_2=50$. 

\begin{figure*}
\centering
\includegraphics[width=.95\textwidth]{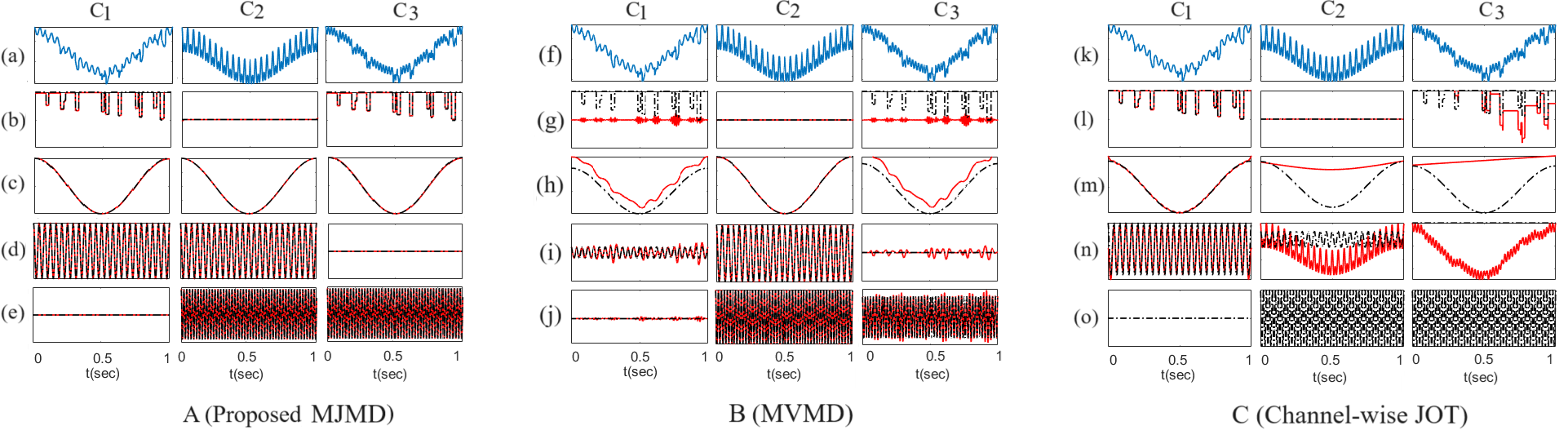}
\caption{Synthetic multivariate signal (rows (a) and (f)), and decomposed components obtained from MJMD method (column A, rows (b) - (e)), MVMD (column B, rows (g) - (j)), and channel-wise JOT (column C, rows (l) - (o)). The red line refers to the extracted components and the black dashed line indicates the original components of the signal in all plots.} 
\label{Synth2}
\end{figure*}

Looking at the above formulation, it can be seen that the jump component only exists in $C_1$, and $C_3$. The result of applying our multivariate JMD scheme is illustrated in Fig. \ref{Synth2} (column A, rows (b) - (e)). 
This figure visualizes the extracted jump and three multivariate oscillatory modes. Our method precisely extracted all components. The comparative results obtained from MVMD are also illustrated in column B. It is clear that MVMD fails to extract jumps as it utilizes a similar basis to VMD which has not been designed to capture jumps. As illustrated in Fig. \ref{Synth2} (rows (g) - (j)), the pure modes are obtained only from channel 2 ($C_2$) where the jump component does not exist. This clearly illustrates that MVMD fails to obtain meaningful components when jump components are present in the signal. Further, channel-wise JOT only precisely obtains the signal components of channel $C_1$ where the channel consists of a jump, a trend, and an oscillation, as shown in Fig. \ref{Synth2} column C. It fails to purely extract all AM-FM oscillatory components for the other channels, resulting in the third oscillatory mode (128 Hz) being absent in channels $C_2$ and $C_3$. Moreover, tuning the input parameters for each channel is cumbersome in channel-wise JOT.

\subsection{Real world signal:}

\subsubsection{ECG-derived respiration}

\begin{figure*}
\centering
\includegraphics[width=01\textwidth]{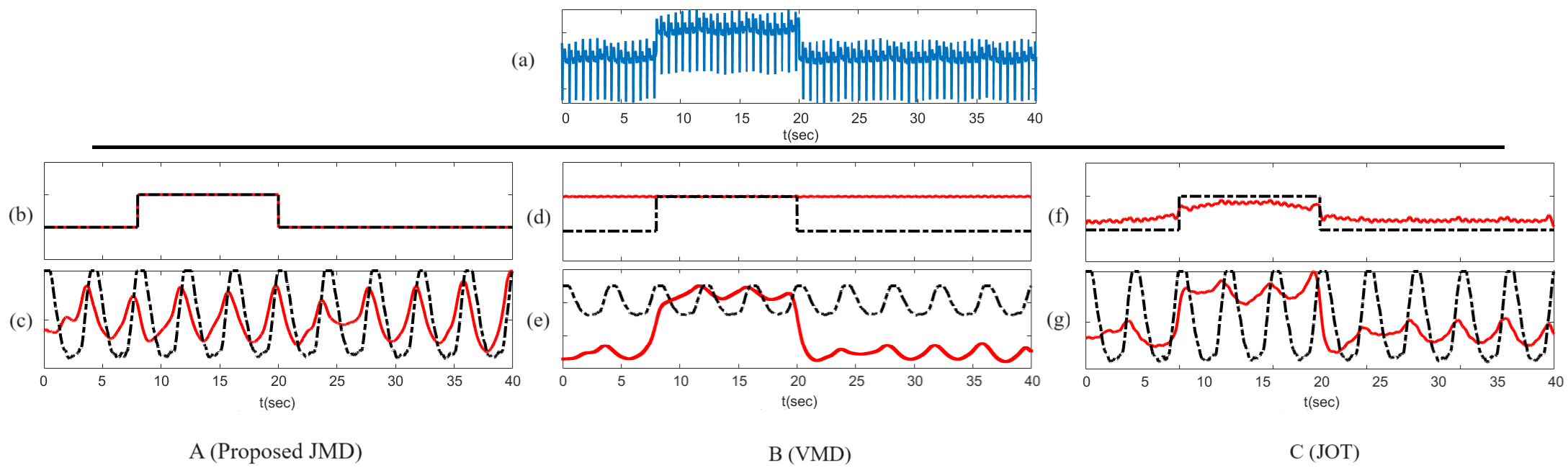}
\caption{Decomposition of the input ECG signal (record number “055m” from MIMIC database) with additive simulated jump component (a) into a jump (b), (d), and (f) and EDR (c), (e), and (g) using our method (Column A), VMD (Column B), and JOT (Column C). The red line refers to the extracted components and the black dashed line indicates the original components of the signal in all plots. In EDR plots (c), (e), and (g), the black dashed line refers to the reference respiratory signal simultaneously recorded with ECG available in the MIMIC dataset.} 
\label{ECG}
\end{figure*}

ECG-derived respiration (EDR) has demonstrated extensive utility in the diagnosis and management of multiple conditions such as stress and sleep disorders in recent decades \cite{nazari2017variational}. Fundamentally, respiration affects the ECG signal by adding a low-frequency component to it. Especially the presence of jumps in ECG, which often manifest as strong harmonics in the low-frequency range (0.3-0.7 Hz), complicates obtaining EDR. The AM-FM signal decomposition methods fail in such scenarios. To assess the efficacy of our method under such conditions, we employ ECG data derived from an initially clean dataset sourced from the MIMIC Database, contaminated with simulated electrode motion artifacts \cite{goldberger2000physiobank}. The simulated motion artifact consists of a jump component characterized by a single rise and a single fall. The signal-to-noise ratio (SNR) measured in dB after adding the artifact to the signal is -6 $dB$. The decomposition results obtained from the JMD method (A), VMD (B), and JOT (C) are depicted in Fig. \ref{ECG}. Notably, we decomposed the input signal into 12 modes using both our method and VMD; however, we only illustrated the jump components ((b) and (d)) and EDR ((c) and (e)) among them. In the case of JOT, we similarly highlighted the components pertinent to EDR and jump. The reference respiratory signal simultaneously recorded with ECG is plotted in rows (c), (e), and (g) to compare the accuracy of all methods. Fig. \ref{ECG} (A) highlights the effectiveness of our proposed approach in obtaining EDR despite the presence of jumps. Comparative results from VMD and JOT indicate that both methods struggle to isolate pure EDR, as the weak low-frequency respiratory signal component is obscured by other harmonics of the wide band and a strong jump in the frequency domain. JOT cannot model AM-FM components, and VMD does not account for the jump component. Obtaining EDR in such a scenario requires a combination of AM-FM signal-jump modeling (i.e., the same formulation as our method) and setting a considerably high $\alpha$.

The input parameters for this experiment were set to $K=13$, $\alpha=2\times10^5$, $\beta=0.2$, $\overbar{b}=0.12$, and $\tau_2=3.6$.

\begin{figure*}[t]
\centering
\includegraphics[width=01\textwidth]{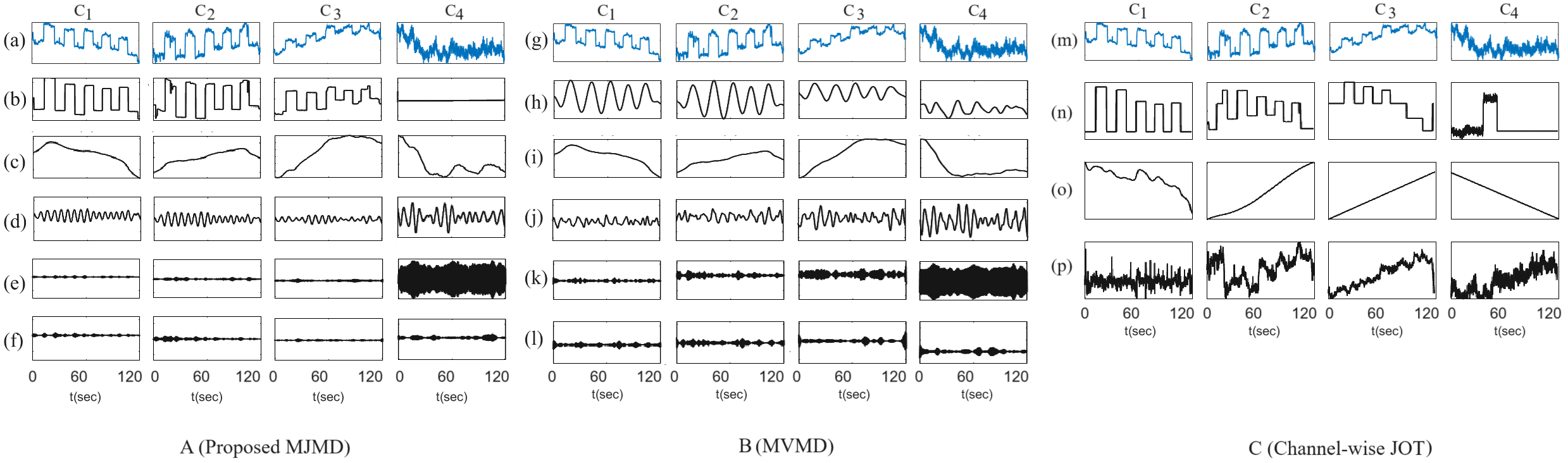}
\caption{Decomposition of the EEG signal (rows (a) and (g)) into a jump (rows (b) and (h)), trend (rows (c) and (i)), alpha-band (rows (e) and (k)), and beta-band (rows (f) and (l)) using our multivariate JMD (MJMD) scheme (column A) and MVMD (column B). Channel-wise JOT (column C) only provides the jump, trend, and a combination of all oscillations.}
\label{EEG}
\end{figure*}

\begin{figure*}[t]
\centering
\includegraphics[width=1\textwidth]{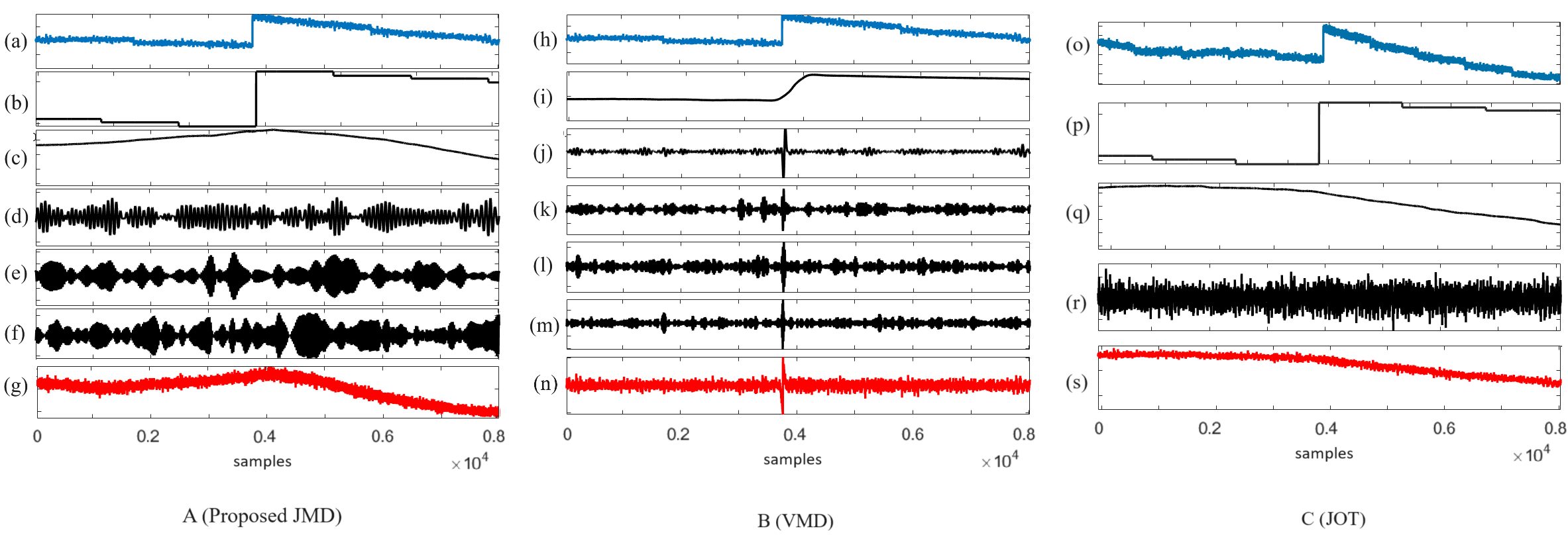}
\caption{Decomposition of the EMF signal (rows (a), (h), and (o)) into jumps, oscillatory components, and trends using our JMD scheme (column A), VMD (column B), and JOT (column C). The last plots of each column (in red) show the reconstructed signals by removing the extracted jump. VMD fails to properly extract the jump with sharp edges, resulting in a spike that destroys the signal quality in the middle where the jump occurred. It also mixes the trend and the jump into a single component (as shown in row (i)).}
\label{EMF}
\end{figure*}

\subsubsection{EEG sub-band time series}

EEG is a multichannel recording of the brain's electrical activity as measured from the scalp. Typically, each channel consists of a time series of electrical signals produced by the brain. EEG signals are known to exhibit oscillatory behavior in specific frequency bands, such as theta (4-8 Hz), alpha (8-12 Hz), beta (13-30 Hz), and gamma (30-150 Hz) bands. Each of these frequency bands is associated with different underlying neural processes. Thus, an essential task in many EEG applications is to break down these neural oscillations into distinct frequency bands.
However, the presence of jumps can significantly impact the decomposition, potentially leading to erroneous and unpure oscillations due to the presence of numerous harmonics across all frequency ranges. This experiment aims to showcase the effectiveness of our proposed approach in identifying and characterizing jumps and inherent frequency modes in motor imagery electroencephalogram (EEG) data. The resulting frequency bands time series hold utility for various applications, including connectivity and power spectrum analysis.

We utilize a publicly available 16-channel EEG dataset collected during motor imagery tasks \cite{cho2017eeg}, and the outcomes (i.e., the results of 4 channels as a few examples) are illustrated in Fig. \ref{EEG}. Data inherently includes huge jumps that extremely affect the results of the conventional AM-FM signal decomposition methods. This example vividly demonstrates our method's capability to capture oscillatory modes despite significant jump interference. Comparative analysis with MVMD and channel-wise JOT highlights their respective limitations in this context: 1) MVMD struggles to isolate jumps and oscillatory modes. The modes obtained from MVMD particularly include jump harmonics across all scales. For instance, the alpha-band modes especially in channels $C_2$ and $C_3$ (Fig. \ref{EEG}, column B, row (k)) show significantly more energy than those obtained from our MJMD method (Fig. \ref{EEG}, column A, row (e)). A similar observation is nearly made for the beta band components (row (l)). This increased energy is likely due to the jump harmonics at this scale, which can lead to incorrect assessments, such as in alpha-band connectivity and power analysis. Additionally, as shown in Fig. \ref{EEG} (column B, row (h)), MVMD mistakenly interprets jumps as oscillations. 2) Channel-wise JOT also fails in obtaining AM-FM multiple frequency bands oscillations. Results obtained from channel-wise JOT (Fig. \ref{EEG}, column C), aside from dealing with the mode mixing problem, demonstrate faulty extracted jumps and trends in channels $C_3$ and $C_4$. One may first use JOT to extract jumps from each channel and then apply the MVMD to extract oscillatory modes associated with each band. This integration requires considerable processing time. 

The input parameters for this experiment were set to $K=15$, $\alpha=2000$, $\beta=0.035$, $\overbar{b}=0.32$, and $\tau_2=3$.

\subsubsection{Earth’s Magnetic and Electric Field (EMF)}

Earth's magnetic and electric fields are actively studied in geophysics for various applications, including atmospheric electricity \cite{lu2016research} and earthquake prediction \cite{yang2016multi}. These signals contain both AM-FM oscillations and abrupt changes. Geomagnetic jerks, which appear as step-like changes, are crucial for interpreting the secular variation and secular acceleration of the magnetic field, thereby providing insights into core dynamics and mantle conductivity \cite{mandea2010geomagnetic}. The oscillatory components are influenced by the solar cycle and its harmonics, annual and semiannual variations, and internal secular variation \cite{courtillot1988time}.

The proposed JMD method effectively extracts both AM-FM modes and step-like abrupt changes, offering a comprehensive analysis of the Earth's magnetic field's temporal behavior. AM-FM mode decomposition methods alone fail to capture meaningful jumps with sharp edges, while methods like JOT only provide pure jump and trend components without decomposing the AM-FM oscillations. Our subsequent analysis will demonstrate the capability of the JMD method to decompose all AF-FM components and jump discontinuities from the signal. However, a detailed analysis of the extracted oscillatory components and other transient phenomena \cite{picozza2019scientific} is beyond the scope of this paper.

In Fig. \ref{EMF} first row, we show a sample of the Earth’s electric field measured by the CSES-01 satellite \cite{diego2020electric}. In the second row, we present the results obtained from our method. We also compare our results with the results obtained from JOT and VMD. The proposed method better separates the jumps producing cleaner oscillatory components. On the contrary, JOT cannot decompose the multiscale oscillatory modes and VMD is not capable of separating the jump component from the oscillations.
Although the JOT framework shows potential to be used for the pre-processing and extraction of trends and jumps which are only two components of this kind of data, our method can be used for the full decomposition of the data into oscillations and jumps in this scenario. This can potentially provide better information about the behavior of the data since the oscillatory components are also available. 

The input parameters for this experiment were set to $K=6$, $\alpha=10^5$, $\beta=0.05$, $\overbar{b}=0.04$, and $\tau_2=8$.

\section{Discussion}
A few comments on the performance and comparison between our JMD and other approaches are in order. First, our (M)JMD are variational approaches capable of decomposing the signal into both oscillatory components and jumps, whereas no other common mode decomposition techniques, including (M)VMD, are not designed to capture jumps. Although discontinuities in the signal can be decomposed using jump extraction methods such as JOT, unlike our JMD, these algorithms do not provide multiple oscillatory components. For instance, JOT decomposes the signals into three parts: jumps, oscillations (including all frequency components), and trends. This means one needs to utilize other decomposition techniques alongside JOT if the extraction of AM-FM modes is also required. In such situations, the decomposition of a signal into jumps and oscillatory modes must be performed in multiple steps, as JOT is naturally a two-step method. Each step in JOT includes an optimization process with its own set of input parameters, which can be challenging to tune. This complexity arises from using individual balancing parameters for each mathematical term in the JOT formulation, concerning each component of the signal. The situation becomes even more cumbersome when additional decomposition techniques such as VMD are required along with JOT to obtain AM-FM modes of the signal. For example, JOT and VMD have five and two key input parameters, respectively. Tuning a total of seven input parameters can be very time-consuming, making our JMD a better solution with only four key input parameters.

Second, the optimization formulations of our (M)JMD are non-convex and therefore not guaranteed to converge. However, there is strong evidence suggesting that the optimization formulations in (M)JMD will converge, as they are based on well-established, non-convex heuristics similar to those used in (M)VMD. Our extensive experiments have shown that our methods converge in all the cases we have considered. Given that (M)VMD has demonstrated empirical convergence properties for a wide range of signals, we expect the (M)JMD algorithms to perform similarly.

Third, regarding the sensitivity of our methods to input parameters, it is important to mention that balancing $K$ and $\alpha$ can be challenging and can significantly affect the decomposition results in our methods. A large $K$ may lead to duplicate modes, while a small $K$ can cause mode mixing problems. Similarly, a high value of $\alpha$ can lead to incorrect noisy modes or cause convergence issues, whereas a low value of $\alpha$ can also result in mode mixing. Essentially, a high value of $\alpha$ is suitable when dealing with a dense spectrum containing frequency components that are very close to each other. Although other input parameters must be set in all these techniques, the methods are not highly sensitive to changes in those parameters. For better convergence of our approaches, we suggest the following suitable ranges for input parameters: (i) $\alpha = 10^3 - 10^5$, (ii) $0 < \beta \le 1$, (iii) $\gamma = 0 - 50$, (iv) $\overline{b} = 0.1 - 0.9$.

\section{Conclusion}

We have presented novel fully data-driven methods for decomposing multivariate signals into jumps and their intrinsic mode functions at multiple frequency scales. By combining tools and methods from traditional AM-FM signal decomposition and jump extraction concepts, we have introduced JMD and MJMD algorithms for obtaining piecewise constant functions (as jumps) and intrinsic oscillatory components of univariate and multivariate signals, respectively.

These methods enable users to model sharp discontinuities in the signal as jumps and separate them from the oscillatory modes. They are based on a variational optimization formulation solved using the alternating direction method of multipliers (ADMM). To our knowledge, no other conventional mode decomposition methods define the signal as composed of wide-band jumps and narrow-band AM-FM signals, incorporating a specific prior to account for jumps in the signal.

Moreover, JMD competes well with other specific jump decomposition methods like JOT, which cannot obtain multiple oscillatory modes of the signals. Additionally, the JOT method is more challenging to initialize compared to our JMD approach, as it requires more input parameters to be predefined due to its use of different priors to account for oscillations and trends separately. In contrast, JMD is more likely to converge since it models trends and oscillations using only one prior.

The proposed methods have been tested on various synthetic and real data, including real-life electroencephalogram (EEG), electrocardiogram (ECG), and Earth's electric field signals, demonstrating the broad applicability of these approaches across different fields.

While the initial promise of the proposed methods has been demonstrated, future work could focus on: i) improving the cost function of the JMD method to better model discontinuities in either the continuous or discrete domain, and ii) eliminating the need for presetting parameter $K$ using a successive approach proposed in \cite{nazari2020successive}.

The MATLAB implementation codes for our JMD and MJMD methods are available at the following link:  \href{https://se.mathworks.com/matlabcentral/profile/authors/16833607}{https://se.mathworks.com/matlabcentral/profile/authors/16833607}.

\section*{Acknowledgment}
Funding: This work was supported by IRRAS USA Inc., San Diego, CA 92130 US.

\bibliographystyle{IEEEtran}
\bibliography{JMD}
\end{document}